\documentclass[aps,prd,fleqn,superscriptaddress]{revtex4}
\usepackage{graphicx,xcolor,natbib,braket}
\usepackage{amsmath,amssymb,amsfonts}
\newcommand{\bse}{\begin{subequations}}
\newcommand{\ese}{\end{subequations}}
\newcommand{\be}{\begin{equation}}
\newcommand{\ee}{\end{equation}}
\newcommand{\bea}{\begin{eqnarray}}
\newcommand{\eea}{\end{eqnarray}}
\newcommand{\ba}{\begin{array}}
\newcommand{\ea}{\end{array}}

\usepackage[colorlinks=true, linkcolor=blue, bookmarks=true]{hyperref}
\begin{document}
%IPM/P-2015/050
\title{On Holographic Time-Like Entanglement Entropy}
\author{ M. M. Daryaei Goki\footnote{$m_{-}$daryaeigoki@sbu.ac.ir}}
\affiliation{Department of Physics, Shahid Beheshti University, 1983969411, Tehran, Iran}
\author{M. Ali-Akbari\footnote{$\rm{m}_{-}$aliakbari@sbu.ac.ir}}
\affiliation{Department of Physics, Shahid Beheshti University, 1983969411, Tehran, Iran}
%\author{A. Davody\footnote{davody@ipm.ir}}
%\affiliation{School of Particles and Accelerators, Institute for Research in Fundamental Sciences (IPM),
%P.O.Box 19395-5531, Tehran, Iran}
%\author{H. Ebrahim\footnote{hebrahim@ut.ac.ir}}
%\affiliation{Department of Physics, University of Tehran, North Karegar Ave., Tehran 14395-547, Iran}
%\affiliation{School of Physics, Institute for Research in Fundamental Sciences (IPM),
%P.O.Box 19395-5531, Tehran, Iran}
%\author{ L. Shahkarami\footnote{l.shahkarami@du.ac.ir}}
%\affiliation{School of Physics, Damghan University, Damghan, 41167-36716, Iran}
\begin{abstract}
In this paper, we compare and analyze holographic timelike entanglement entropy with pseudo-entropy in quantum mechanics for a two-qubit system, considering transitions from a thermal state to an anisotropic thermal state at fixed temperature. We identify several common properties but also report important differences between these quantities.   
\end{abstract}

\maketitle
\tableofcontents

\section{Introduction}
The AdS/CFT correspondence, or more generally gauge-gravity duality, establishes a profound equivalence between a gravitational theory in a $(d+1)$-dimensional (asymptotically) Anti-de Sitter (AdS) background and a conformal field theory (CFT) defined on the $d$-dimensional boundary of the geometry \cite{Maldacena:1997re}. In this framework, particularly in low-energy limits, physical phenomena in the strongly coupled quantum field theory can be described by classical gravitational dynamics in the dual bulk geometry \cite{Witten:1998qj}. This correspondence enables the computation of various quantum quantities through geometric methods. Consequently, this duality has become a central paradigm in modern theoretical physics, providing deep insights into quantum gravity \cite{Aharony:1999ti} and allowing calculation of quantities such as entanglement entropy \cite{Ryu:2006bv, Ryu:2006ef} and pseudo-entropy \cite{Caputa:2024gve}.

Entanglement entropy, as a non-local observable, is a fundamental quantity in quantum information and quantum field theory that quantifies the degree of quantum correlations between two subsystems \cite{Calabrese:2004eu}. When a system is partitioned into a subregion $A$ and its complement $\bar{A}$, the entanglement entropy measures how much information about one part is encoded in the other \cite{Casini:2009sr}. It is defined as the von Neumann entropy of the reduced density matrix $\rho_A$ \cite{Nielsen:2012yss}
\begin{equation}
S_{eA} = - \mathrm{Tr}_A (\rho_A \log \rho_A),\qquad
\rho_A = \mathrm{Tr}_{\bar{A}} \, |\psi\rangle \langle \psi |, 
\end{equation}
where $|\psi\rangle$ is a pure state describing the physical system and the reduced density matrix is obtained by tracing over the degrees of freedom in $\bar{A}$. The entanglement entropy $S_{eA}$ is an intrinsic property of quantum theory and thus has no classical counterpart.

Although calculating entanglement entropy is generally difficult, it admits a remarkably simple geometric description within gauge-gravity duality. Ryu and Takayanagi first proposed in \cite{Nishioka:2009un} that entanglement entropy can be computed holographically as
\begin{equation}\label{prescription}
S_{eA} = \frac{\mathrm{Area}(\Gamma_A)}{4G_N^{(d+2)}},
\end{equation}
where $G_N^{(d+2)}$ is the $(d+2)$-dimensional Newton constant and $\Gamma_A$ is a codimension-2 minimal surface whose boundary coincides with the boundary of the subregion $A$ on the asymptotic boundary of the bulk, {\it{i.e.}} $\partial\Gamma_A = \partial A$. This proposal has attracted considerable attention over the past decade and has passed several non-trivial tests against known quantum field theory results. For further details, we refer the reader to \cite{Ling:2015dma,Klebanov:2007ws,Liu:2013iza,VanRaamsdonk:2010pw,Nishioka:2018khk}.

To introduce pseudo-entropy as a generalization of entanglement entropy, consider two arbitrary states $\ket{\psi}$ and $\ket{\phi}$ in the full Hilbert space $\mathcal{H}$, which decomposes as $\mathcal{H}_A\otimes\mathcal{H}_{\bar{A}}$. The transition matrix is defined as 
\begin{equation}
\mathcal{T}=\frac{\ket{\psi}\bra{\phi}}{\braket{\phi|\psi}},
\end{equation}
which is generally non-Hermitian \cite{Mollabashi:2021xsd}. The reduced transition matrix for region $A$ is then
\begin{equation}
\mathcal{T}_A=\mathrm{Tr}_{\bar{A}}(\mathcal{T}),
\end{equation}
and the pseudo-entropy is given by 
\begin{equation}\label{peso}
S_{pA}=-\mathrm{Tr}_{A}(\mathcal{T}_A\log\mathcal{T}_A).
\end{equation} 
Since $\mathcal{T}$ is non-Hermitian, its eigenvalues can be complex and consequently the pseudo-entropy may take complex values. Note that when $\ket{\psi}=\ket{\phi}$, pseudo-entropy clearly reduces to entanglement entropy. Intuitively, pseudo-entropy can be interpreted as a measure of how much relationship exists between different states in the Hilbert space as accessed through region $A$.

Takayanagi and collaborators have introduced timelike entanglement entropy as an extension of entanglement entropy to timelike-separated regions \cite{Doi:2022iyj,Doi:2023zaf}. The timelike entanglement entropy contains imaginary part and therefore is a pseudo-entropy rather than a von Neumann entropy. In the holographic description, {\it{i.e.}} holographic time-like entanglement entropy (HTEE), the associated extremal surfaces, $\Gamma_A$s in \eqref{prescription}, are not constrained to be spacelike and can become timelike. This framework provides a way to probe quantum correlations across timelike separations. While the real part of timelike entanglement entropy reduces to the usual entanglement entropy in appropriate limits, the appearance of an imaginary contribution represents a genuinely new feature.
The imaginary part of timelike entanglement entropy does not admit a direct interpretation as information loss. Instead, it is sensitive to relative phases and interference effects in transition amplitudes, suggesting a natural connection to quantum coherence \cite{Nakata:2020luh}.
Motivated by this phase-sensitive structure, it has been suggested that the imaginary part of timelike entanglement entropy may offer a perspective on the notion of time \cite{Takayanagi:2025ula}. Since coherence and phase relations play a central role in temporal ordering and quantum evolution, aspects of time may be understood as emerging from such phase-dependent correlations with time appearing as an effective rather than fundamental concept.

In what follows, we compute pseudo-entropy in quantum mechanics for transitions from an initial state to an anisotropic thermal state and analyze its properties. We then calculate the HTEE in an anisotropic background and compare the results with those from quantum mechanics. The question we address is whether the behavior of HTEE, compared to pseudo-entropy, can serve as a diagnostic to distinguish strongly coupled field theories. In other words, the response of HTEE to anisotropy remains largely unexplored and may reveal distinctive features of strongly coupled systems.

\section{Pseudo-entropy in quantum mechanics: A toy model}\label{example}
As a warm-up, in this section we attempt to compute the pseudo-entropy for two thermal states in the presence of anisotropy. For this purpose, we take the initial and final states to be
\begin{equation}\label{isotropic}
\rho_i= \frac{e^{-\beta H_i}}{\operatorname{Tr}(e^{-\beta H_i})}, \qquad \text{where} \qquad H_i = \sigma_1^x \sigma_2^x + \sigma_1^y \sigma_2^y + \sigma_1^z \sigma_2^z,
\end{equation}
and
\begin{equation}\label{final}
\rho_f= \frac{e^{-\beta H_f}}{\operatorname{Tr}(e^{-\beta H_f})}, \qquad \text{where} \qquad H_f = \Delta(\sigma_1^x \sigma_2^x + \sigma_1^y \sigma_2^y) + \delta\, \sigma_1^z \sigma_2^z.
\end{equation}
Several remarks are in order. First, the $\sigma$'s denote Pauli matrices and the indices $1$ and $2$ label the two qubits of the system. Second, both mixed states share the same inverse temperature $\beta = T^{-1}$. Third, the appearance of the coefficients $\Delta$ and $\delta$ in $H_f$ introduces anisotropy into the final state; the transition between $\rho_i$ and $\rho_f$ can therefore be viewed as a symmetry-breaking process. Finally, the parameters $\Delta$ and $\delta$ are taken to be complex.

\begin{figure}
%\centering
\includegraphics[width=88 mm]{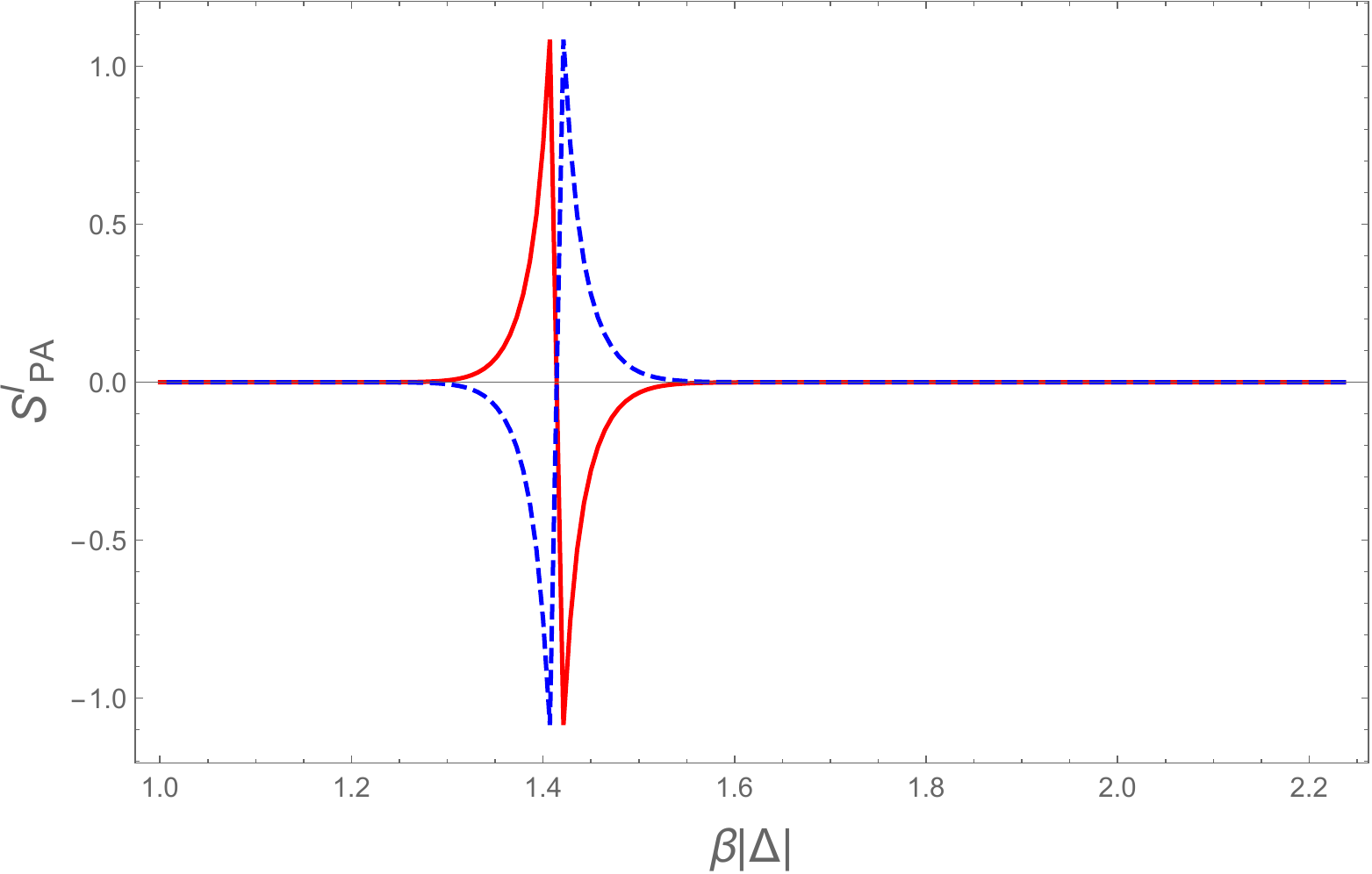}  
\includegraphics[width=88 mm]{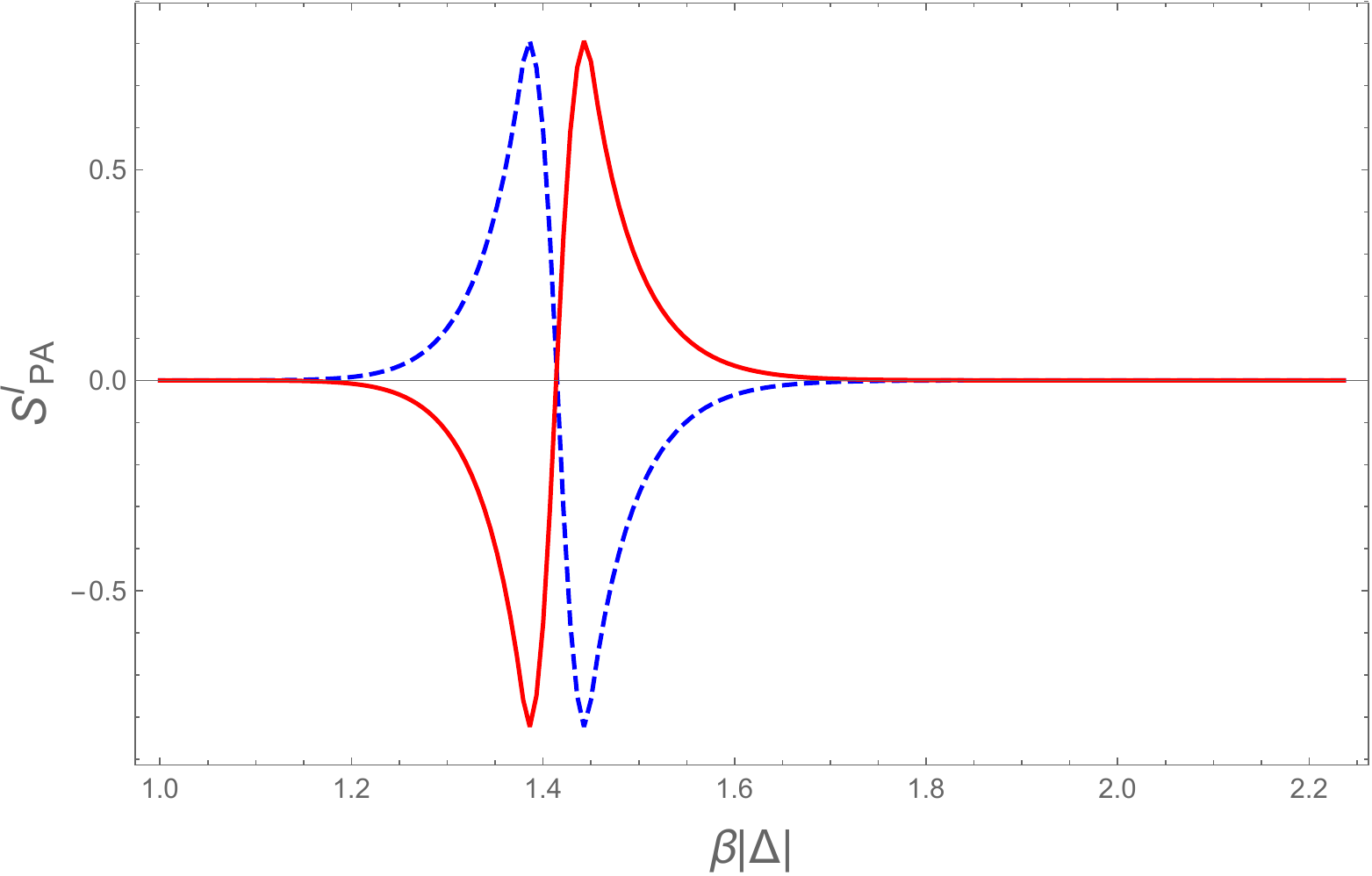}   
\caption{The imaginary part of the pseudo-entropy $S_{pA}^I$ is plotted as a function of $\beta\vert\Delta\vert$ for $\beta=10$ and $0\le\Delta_x\le 2$, left panel: $\ \delta_x=1,\ \delta_y=0$ and $\Delta_y=1$ (red) or $\Delta_y=-1$ (blue, dashed) and right panel: $\ \delta_x=-1,\ \delta_y=1.1$ and $\Delta_y=1$ (red) or $\Delta_y=-1$ (blue, dashed). Two opposite behaviors arise depending on the sign of the anisotropy parameter $\Delta_y$.}
\label{fig1}
\end{figure}

Now, after a lengthy but straightforward calculation, the transition matrix is found to be
\begin{align}
\mathcal{T} % \frac{\rho_i \rho_f}{{\text{Tr}}(\rho_i \rho_f)}, 
=\frac{1}{D}\rm{diag}\left(e^{-\beta(1+\delta)}, e^{-\beta(-\delta+2\Delta-3)}, e^{-\beta(-\delta-2\Delta+1)}, e^{-\beta(1+\delta)}\right),
\end{align}
where $D=\left(2e^{-\beta(1+\delta)}+e^{-\beta(-\delta+2\Delta-3)}+e^{-\beta(-\delta-2\Delta+1)}\right)$. 
Tracing out the second qubit yields the reduced transition matrix  $\mathcal{T}_A$  for subsystem $A$
\begin{equation}
\mathcal{T}_A ={\text{Tr}}_{\bar{A}}(\mathcal{T})=\frac{1}{D}\rm{diag}\left(\lambda_1,\lambda_2\right),
\end{equation}
where
\begin{align}
\lambda_1&= e^{-\beta(1+\delta)}+e^{-\beta(-\delta+2\Delta-3)},\\
\lambda_2&= e^{-\beta(1+\delta)}+e^{-\beta(-\delta-2\Delta+1)}.
\end{align}
To extract the real and imaginary parts of the pseudo-entropy, we decompose the complex parameters as
\begin{align}
\Delta&= \Delta_x+i \Delta_y\\
\delta&= \delta_x+i \delta_y, 
\end{align} 
and the real and imaginary parts of $\lambda$s are then 
\bse\begin{align}
\lambda_{1R}&=\frac{c_1 c_3+c_2 c_4}{c_3^2+c_4^2},\ \ \ \ \lambda_{1I}=\frac{c_2 c_3-c_1c_4}{c_3^2+c_4^2},\\
\lambda_{2R}&=\dfrac{c_5 c_3+c_6 c_4}{c_3^2+c_4^2},\ \ \ \ \ \lambda_{2I}=\frac{c_6 c_3-c_5 c_4}{c_3^2+c_4^2},
\end{align}\ese
where
\bse
\begin{align}
c_1&=e^{-\beta(1+2\delta_x)}\cos(2\beta\delta_y)+e^{-\beta(2\Delta_x-3)}\cos(2\beta\Delta_y),\\ 
\label{c2} c_2&=-e^{-\beta(1+2\delta_x)}\sin(2\beta\delta_y)-e^{-\beta(2\Delta_x-3)}\sin(2\beta\Delta_y),\\
c_3&=2e^{-\beta(1+2\delta_x)}\cos(2\beta\delta_y)+e^{-\beta(2\Delta_x-3)}\cos(2\beta\Delta_y)+e^{-\beta(1-2\Delta_x)}\cos(2\beta\Delta_y),\\
\label{c4} c_4&=-2e^{-\beta(1+2\delta_x)}\sin(2\beta\delta_y)-e^{-\beta(2\Delta_x-3)}\sin(2\beta\Delta_y)+e^{-\beta(1-2\Delta_x)}\sin(2\beta\Delta_y),\\
c_5&=e^{-\beta(1+2\delta_x)}\cos(2\beta\delta_y)+e^{-\beta(1-2\Delta_x)}\cos(2\beta\Delta_y),\\
\label{c6} c_6&=-e^{-\beta(1+2\delta_x)}\sin(2\beta\delta_y)+e^{-\beta(1-2\Delta_x)}\sin(2\beta\Delta_y).
\end{align}
\ese
Writing $\lambda_i = r_i e^{i\theta_i}$ and using \eqref{peso}, the pseudo-entropy $S_{pA}$ decomposes into real and imaginary parts as
\bse\begin{align}
S_{pA}&=S_{pA}^R+i S_{pA}^I,\\
S_{pA}^R&=-\lambda_{1R}\log r_1+\lambda_{1I}\theta_1-\lambda_{2R}\log r_2+\lambda_{2I}\theta_2,\\
S_{pA}^I&=-\lambda_{1I}\log r_1-\lambda_{1R}\theta_1-\lambda_{2I}\log r_2-\lambda_{2R}\theta_2,
\end{align}\ese
where
\begin{align}
r_1&=\sqrt{\lambda_{1R}^2+\lambda_{1I}^2},\ \ \ \ \theta_1=\tan^{-1}(\dfrac{\lambda_{1I}}{\lambda_{1R}}),\\
r_2&=\sqrt{\lambda_{2R}^2+\lambda_{2I}^2},\ \ \ \ \theta_2=\tan^{-1}(\dfrac{\lambda_{2I}}{\lambda_{2R}}).
\end{align}
Our final results show a few important points as follows:
\begin{itemize}
\item  The imaginary part of the pseudo-entropy, $S_{pA}^I$, vanishes when the $\lambda_i$ are real, \textit{i.e.} when $\Delta_y = \delta_y = 0$. This special case corresponds to taking both parameters $\Delta$ and $\delta$ as real numbers. Consequently, to examine the influence of anisotropy on pseudo-entropy, it is necessary to choose at least one of these parameters to be complex. This implies that the reduced transition matrix $\mathcal{T}_A$ must possess at least one complex eigenvalue. Therefore, note that the imaginary part of the pseudo-entropy is zero in the isotropic case \eqref{isotropic} even in the presence of temperature.

\item In the high-temperature limit, $\beta \to 0$, {\it{i.e.}} when the temperature is large compared to all other scales in the theory, such as the anisotropy parameters, \eqref{c2}, \eqref{c4} and \eqref{c6} show clearly that the imaginary parts of $\lambda_1$ and $\lambda_2$ vanish. Consequently, the imaginary part of the pseudo-entropy also vanishes in this limit. Note that, irrespective of the values of $\Delta$ and $\delta$, $S_{pA}^I$ is zero at high temperature. In other words, for the present example, a non-zero $\beta$ is necessary but not sufficient to obtain a non-zero imaginary part of the pseudo-entropy.

\item To compare our results with those from holography, it would be natural to choose $\Delta_x = 1$, $\Delta_y = 0$ and arbitrary $\delta$ and $\beta$. However, this corresponds to a special case for which the imaginary part of $S_{pA}$ vanishes. Therefore, in the following analysis, we fix $\delta$ and $\beta$ and plot the results as functions of $\Delta$.

\item In the low-temperature limit $\beta \to \infty$, \eqref{c2}, \eqref{c4} and \eqref{c6} show that for $\delta_x > -0.5$ and $\Delta_x > 1.5$, both components of the pseudo-entropy vanish. By contrast, in the region $\delta_x < -0.5$ and $\Delta_x < 1.5$, certain parameter choices could in principle yield a non-zero imaginary part.

\item To compare the results obtained from the holographic approach with those coming from the present model, we plot the imaginary part of the pseudo-entropy as a function of the dimensionless parameter $\beta|\Delta|$ where $|\Delta|=\sqrt{\Delta_x^2+\Delta_y^2}$ for selected values of the anisotropy parameters $\Delta$ and $\delta$ in figure \ref{fig1}. While many different plots could be produced for arbitrary choices of these parameters, we show only two representative cases as illustrations. In figure \ref{fig1}, the imaginary part of the pseudo-entropy is plotted as a function of $\beta|\Delta|$ and two distinct behaviors are observed.

\item In the quantum mechanical calculation, the coupling strength (strong or weak) plays no direct role. Once the eigenstates and eigenvalues of the initial and final Hamiltonians are specified, the results can be computed. In other words, the states $\ket{\psi}$ and $\ket{\phi}$ may correspond to either strongly or weakly coupled systems.
\end{itemize}

\section{Review on the background}
We consider a background geometry described by the following 5-dimensional metric:
\begin{align}
ds^{2}=f_1(u)\,dt^2+f_2(u)\,du^2+f_3(u)\,dz^2+f_4(u)\,(dx^2+dy^2),
\label{metric}
\end{align}
where $u$ is the radial coordinate and the boundary is located at $u=0$. This background asymptotically approches AdS$_5\times$S$^5$ metric and the functions $f_1(u),\dots,f_4(u)$ depend only on radial coordinate. Here, $f_1$ and $f_2$ are blackening factors that vanish at the horizon position $u=u_h$, {\it{i.e.}}
\be
f_1(u_h)=0, \qquad f_2^{-1}(u_h)=0.
\ee
The metric clearly exhibits an anisotropy between the $z$ direction and the $xy$-plane whenever $f_3(u) \neq f_4(u)$. An isotropic background is recovered when these two metric components are set equal.
 
A particularly interesting anisotropic background, which corresponds to pressure anisotropy in quark--gluon plasma, was introduced in \cite{Casalderrey-Solana:2011dxg}. Then the metric is given by 
\begin{align}
f_1(u)= -\mathcal{F} \mathcal{B}u^{-2},\quad f_2(u)= \mathcal{F}^{-1}u^{-2},\quad f_3(u)= \mathcal{H}u^{-2},\quad f_4(u)=u^{-2}.
\label{10}
\end{align}
The functions $\mathcal{H}$, $\mathcal{F}$ and $\mathcal{B}$ depend only on the radial direction and they are given by the following expressions in terms of the dilaton field $\phi$
\begin{align}
\mathcal{H}&=e^{\phi}, \\
\label{eomf} \mathcal{F}&=\dfrac{e^{-\frac{1}{2}\phi}(a^2 e^{\frac{7}{2}\phi}(4u+u^2\phi')+16\phi')}{4(\phi'+u\phi'')},\\
\label{eomb}\dfrac{\mathcal{B}'}{\mathcal{B}}&=\dfrac{1}{24+10u\phi'}(24\phi'-9u\phi'^2+20u\phi''),
\end{align}
where the dilaton field satisfies the following third-order equation
\begin{align}\begin{split}
&\dfrac{256\phi'\phi''-16\phi'^3(7u\phi'+32)}{ua^2e^{\frac{7}{2}\phi}(u\phi'+4)+16\phi'}+\dfrac{\phi'}{u(5u\phi'+12)(u\phi''+\phi')}\times[13u^3\phi'^4+8u(11u^2\phi''^2-60\phi''-12u\phi''')\\&+u^2\phi'^3(13u^2\phi''+96)+2u\phi'^2(-5u^3\phi'''+53u^2\phi''+36)+\phi'(30u^4\phi''^2-64u^3\phi'''-288+32u^2\phi'')]=0.
\end{split}
\label{eomphi}
\end{align}
In order to solve the above equation of motion, one needs the suitable boundary conditions. The boundary of the geometry \eqref{metric}, where the filed theory lives, is located at $u=0$ and the boundary conditions are 
\be\label{bc}
\phi_0=0,\quad \mathcal{H}_0=\mathcal{F}_0=1.
\ee
where $\phi_0\equiv\phi(u=0)$ and so on. Since the dilaton field equation of motion is a third-order differential equation, solving it requires two initial conditions. To specify these conditions conveniently, we define the shifted field $\tilde{\phi}(u)\equiv\phi(u) +\frac{4}{7}\log a$.
This redefinition removes the explicit dependence on $a$ in \eqref{eomphi} and produces an overall factor of $a^{2/7}$ in \eqref{eomf}. One can then expand the dilaton field near the horizon as
\begin{equation}\label{extend}
 \tilde{\phi}(u)=\tilde{\phi}_h+\sum_{n\geqslant1}\tilde{\phi}_n(u_h)\,(u-u_h)^n,
\end{equation}
where \(\tilde{\phi}_h\equiv\tilde{\phi}(u_h)\). Using \eqref{eomphi}, the first two coefficients in \eqref{extend} can be found.
For given values of $u_h$ and $\tilde{\phi}_h$, these initial conditions together with the boundary condition \eqref{bc} allow us to solve the dilaton equation of motion numerically. The remaining metric components are then determined using \eqref{eomf} and \eqref{eomb}. Thus, the full solution is characterized by two parameters: the value of the dilaton field at the horizon and the horizon location itself. Finally, the temperature and entropy density of the solution are obtained as
\begin{align}\begin{split}
&T=-\dfrac{1}{4\pi}\mathcal{F}'(u_h)\sqrt{\mathcal{B}(u_h)},\\&
\frac{s}{N_c^2}=\dfrac{1}{2}\dfrac{e^{-\frac{5}{4}\phi_h}}{\pi u_h^3}
\end{split}
\label{15}
\end{align}
where $N_c$ is a constant.  For more details, we refer the reader to the original paper \cite{Mateos:2011tv}. 

In the case of high temperature limit, ${\it{i.e.}}\ aT^{-1}\ll 1$, an analytic solution was introduced as follows
\begin{align}\begin{split}
&\mathcal{F}=1-\dfrac{u^4}{u_h^4}+a^2\widehat{\mathcal{F}}_2(u),\\&
\mathcal{B}=1+a^2 \widehat{\mathcal{B}}_2(u),\\&
\phi=a^2\widehat{\phi}_2(u),
\end{split}
\label{16}
\end{align}
where
\begin{align}\begin{split}
&\widehat{\mathcal{F}}_2(u)=\dfrac{1}{24u_h^2}(8u^2(u_h^2-u^2)-10u^4\log 2+(3u_h^4+7u^4)\log(1+\dfrac{u^2}{u_h^2})),\\&
\widehat{\mathcal{B}}_2(u)=-\dfrac{u_h^2}{24}(\dfrac{10u^2}{u_h^2+u^2}+\log(1+\dfrac{u^2}{u_h^2})),\\&
\widehat{\phi}_2(u)=-\dfrac{u_h^2}{4}(\log(1+\dfrac{u^2}{u_h^2})).
\end{split}
\label{17}
\end{align}
The temperature and entropy density is then obtained as
\begin{align}\begin{split}
&T=\dfrac{1}{\pi u_h}+\dfrac{(5\log2-2)u_h}{48\pi^2}a^2,\\
&\frac{s}{N_c^2}=\dfrac{\pi^2T^3}{2}+\dfrac{T}{16}a^2.
\end{split}
\end{align}

It is important to note that this background has been studied from various perspectives. Previous work has shown that physical quantities, such as quasi-normal modes, depend on the dimensionless combination $aT^{-1}$ rather than separately on $a$ and $T$, since the underlying theory is conformal \cite{Ali-Akbari:2014xea, Rahimi:2018ica}.
In the next section, we use this anisotropic metric to calculate the HTEE.

\section{Holographic timeLike entanglement entropy }
\begin{figure}
\centering
\includegraphics[width=80 mm]{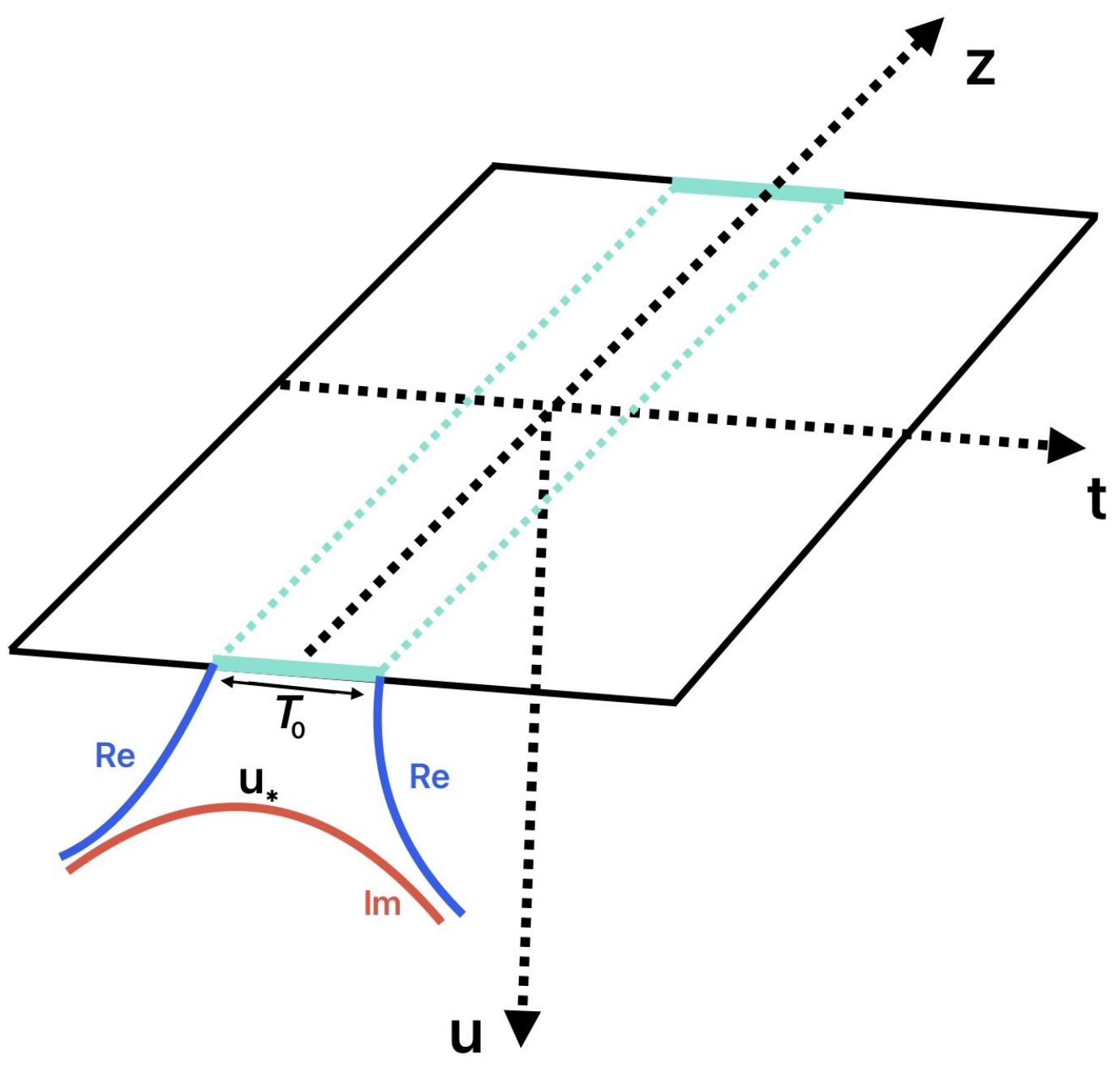}   
\includegraphics[width=90 mm]{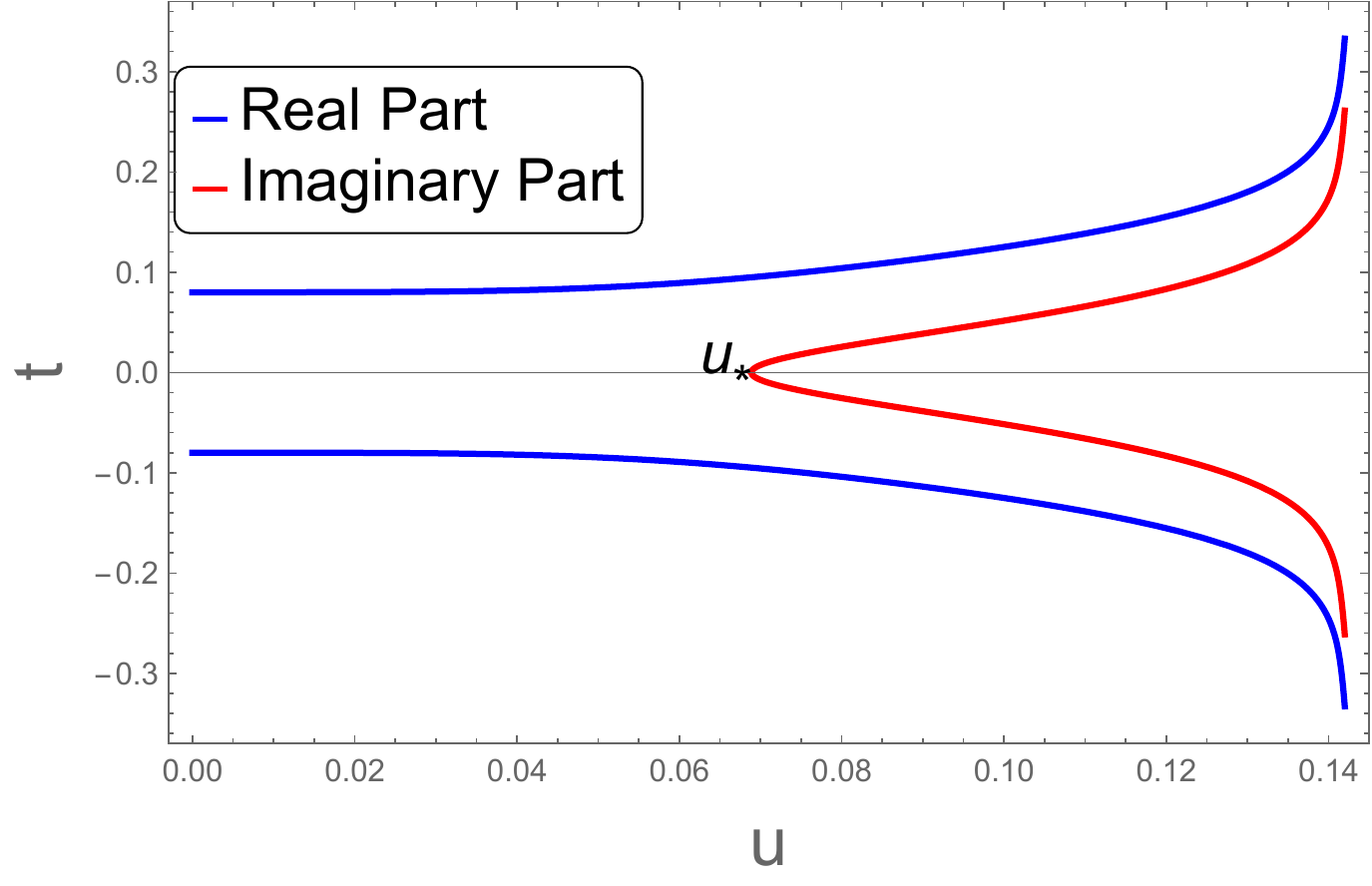}   
\caption{The real and imaginary parts of the HTEE are shown for $T_0=0.16$, $T=2.30$, $a=7.674$ and $c=10^6$.}
\label{fig1a}
\end{figure}

The background introduced in \eqref{metric} is holographically dual to an anisotropic strongly coupled quantum field theory \cite{Mateos:2011tv}. To compute the HTEE and study the effect of anisotropy, we consider an infinite strip-like subsystem $A$ given by
\begin{equation}
A = \left\{ \begin{array}{ll}
-\dfrac{T_0}{2} \leq t \leq \dfrac{T_0}{2}, & \\[2mm]
y = 0, &
\end{array} \right.
\end{equation}
where $T_0$ denotes the length of the time interval, see figure \ref{fig1a}, left panel. Owing to the $U(1)$ symmetry between the $x$ and $y$ directions, the choice of subsystem $A$ along either axis {\it{i.e.}} $x$ or $y$ is equivalent.
Following the holographic prescription \eqref{prescription} and parameterizing the minimal surface by $t(u)$, the HTEE is obtained by extremizing the area integral
\begin{align}
S_t = \frac{V_{3}}{4G_N^{(5)}} \int_{u_i}^{u_f} du \, \sqrt{f_3 f_4 \bigl( f_1 t'^2 + f_2 \bigr)},
\label{4}
\end{align}
where $t' \equiv dt/du$, and $V_{3}$ is the volume of the transverse coordinates.
Treating the integrand in \eqref{4} as a Lagrangian, the equation of motion for $t(u)$ follows as
\begin{align}
t'^2 = \frac{c^2 f_2}{f_1 \bigl( f_1 f_3 f_4 - c^2 \bigr)},
\label{55}
\end{align}
with $c$ an integration constant. The constant $c^2$ can be expressed as $\pm C^2$ where $C^2 = -f_{1*} f_{3*} f_{4*}$ which is always positive. Here $f_{i*} \equiv f_i(u_*)$ and $u_*$ is defined by $t'(u_*) \rightarrow \infty$ which by symmetry lies at $t = 0$.

Now, substituting \eqref{55} into \eqref{4} and decomposing $S_t$ into the real and imaginary part as 
\be 
S_t=S_{tR}+i S_{tI},
\ee
we obtain 
\begin{itemize}
\item {\bf{Case with the plus sign: Space-like surface}}\\
In this case, \eqref{55} has no turning point, equivalently, $t'(u) \neq 0$ everywhere except at the boundary. This yields a purely real HTEE. The blue curve in the right panel of Fig.~\ref{fig1a} shows $S_{tR}$. The integration extends from the horizon to the boundary:
\begin{align}
S_{tR} = \frac{V_{3}}{4G_N^{(5)}} \int_{\epsilon}^{u_h} du \, f_3 f_4 \sqrt{\frac{f_2 f_1}{f_1 f_3 f_4 + f_{1*}f_{3*}f_{4*}}},
\label{6}
\end{align}
where $\epsilon$ is a UV cutoff. The appropriate boundary condition for solving \eqref{55} is $t(\epsilon) = \pm T_0/2$. Due to the UV cutoff, $S_{tR}$ is divergent and must be regularized.

\item{\bf{Case with the minus sign: Time-like surface}}\\
Here, \eqref{55} admits a turning point $u_*$, so the integration domain extends from $u_*$ to the horizon. Since $|f_{1*}|f_{3*}f_{4*} > |f_1|f_3f_4$, this yields the imaginary part of the HTEE:
\begin{align}
S_{tI} = \frac{V_{3}}{4G_N^{(5)}} \int_{u_*}^{u_h} du \, f_3 f_4 \sqrt{\frac{f_2 f_1}{f_1 f_3 f_4 - f_{1*}f_{3*}f_{4*}}}.
\label{5}
\end{align}
As an illustration, the red curve in the right panel of Fig.~\ref{fig1a} shows $S_{tI}$. The corresponding boundary condition is $t'(u_*) \to \infty$.
\end{itemize}

To obtain the regularized real part of the HTEE, we work with an isotropic black hole background of the form
\begin{align}
-u^2 f_1(u)=u^{-2}f^{-1}_2(u)=1-\frac{u^4}{u_h^4},\qquad f_3(u)=f_4(u)=u^{-2}.
\label{7}
\end{align}
Within this setup, the isotropic components of the HTEE, denoted $S^0_{tR}$ and $S^0_{tI}$, can be computed directly. While the imaginary part is already finite, both components are regularized in a consistent manner to yield well-defined results:
\begin{align}
\widehat{S}_{tR} &\equiv \frac{4G^{(5)}_N}{V_{3}}\left( S_{tR} - S_{tR}^0 \right), \\
\widehat{S}_{tI} &\equiv \frac{4G^{(5)}_N}{V_{3}}\left( S_{tI} - S_{tI}^0 \right).
\label{eq:regularized_HTEE}
\end{align}
Here $\widehat{S}_{tR}$ and $\widehat{S}_{tI}$ represent the finite, renormalized quantities. It is worth noting that, in the appropriate limit, the real part of the pseudo-entropy reduces to ordinary entanglement entropy, which has been studied extensively in earlier work~\cite{Afrasiar:2024ldn,Nunez:2025ppd,Gong:2025pnu,Nunez:2025puk,Zhao:2025zgm}. Therefore, the present analysis focuses on the less-explored imaginary part.
Note that in our regularization scheme, $S_{tR(I)}$ and $S_{tR(I)}^0$ are evaluated and subtracted at fixed temperature (not at fixed horizon position $u_h$) so that the influence of the anisotropy parameter is cleanly isolated. In the following section, we employ these definitions to present our numerical findings.

An important point to emphasize is that the final state \eqref{final} is characterized by the temperature together with the parameters $\delta$ and $\Delta$. Fixing the temperature, which is the same for both initial and final states, also determines the initial state. Correspondingly, on the gravity side, our final state \eqref{10} is described by $T$ and $a$, while the initial state at the same temperature is given by \eqref{7}.

\section{Numerical results}

\begin{figure}
%\centering
\includegraphics[width=88 mm]{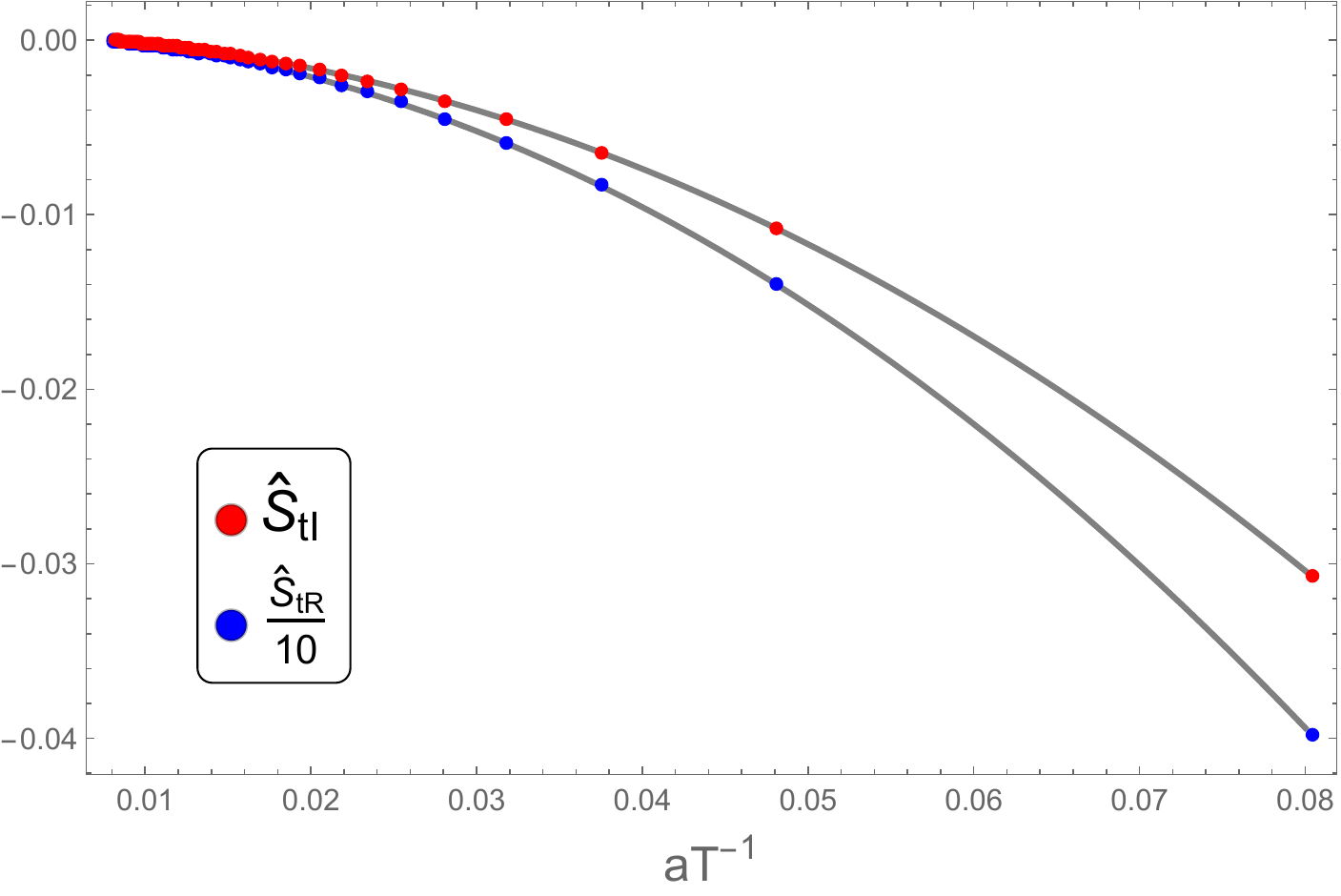}  
\includegraphics[width=85 mm]{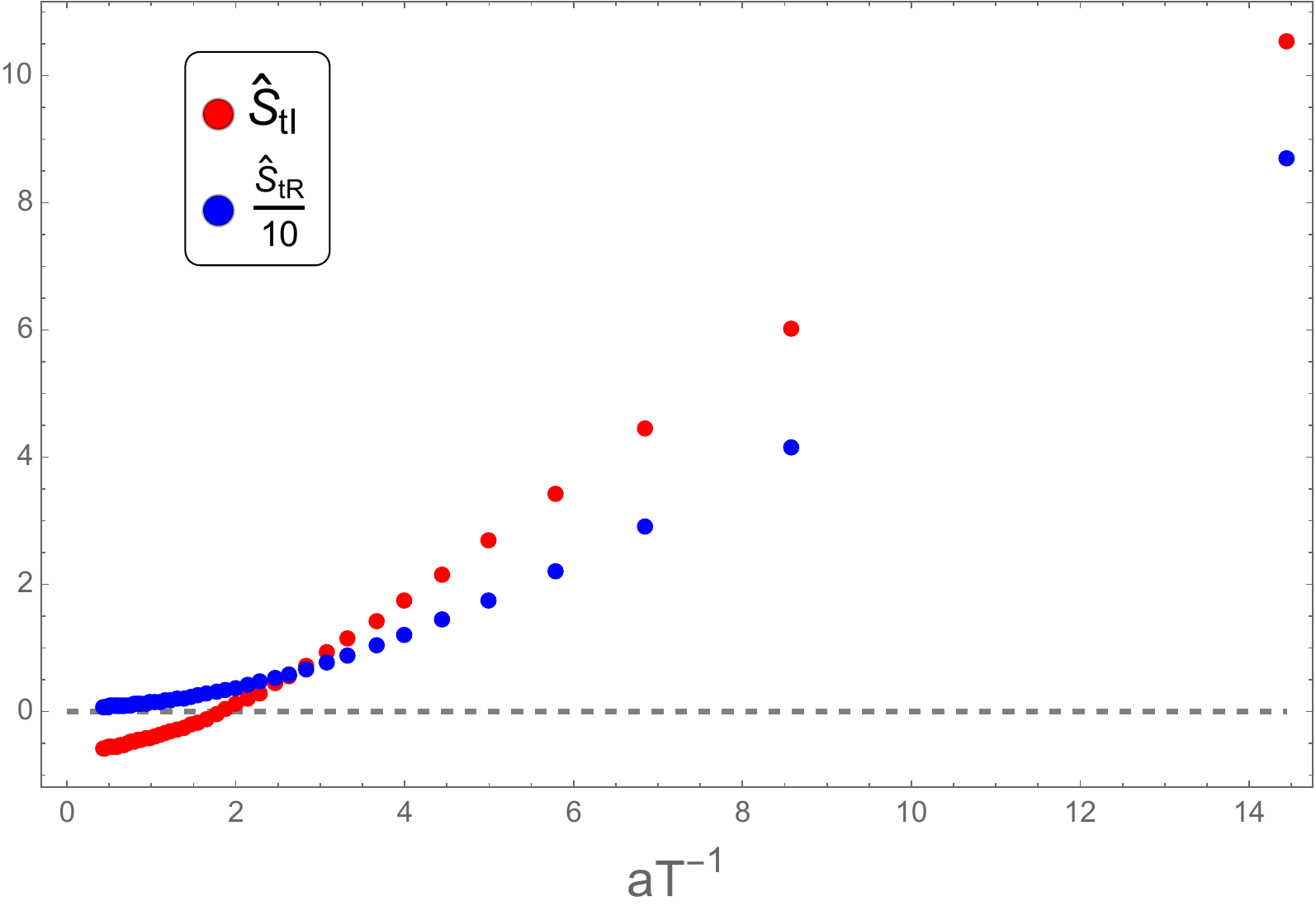}   
\caption{The real and imaginary parts of the HTEE are shown for $T_0 = 0.08$ and $c=10^7$, Left: $T = 10$ and Right: $T=2.3$. In the left panel, $aT^{-1} \ll 1$, while in the right panel, $a$ is arbitrary. The continuous curves correspond to the fitted functions $\widehat{S}_{tR} \propto a^2$ and $\widehat{S}_{tI} \propto a^2$.}
\label{fig2}
\end{figure}

In the left panel of figure \ref{fig2}, the real and imaginary parts of the regularized HTEE are plotted as functions of the anisotropy parameter $a$ at fixed and high temperature, using the background metric \eqref{metric} and \eqref{16}. As is evident from the figure, both the real and imaginary parts of the regularized HTEE decrease as $a$ increases. Consistent with the functional form of the background, the numerical results for both parts are fitted with a second-order polynomial in $a$. In the limit $aT^{-1} \ll 1$ at fixed temperature, the presence of anisotropy reduces the values of both components of the regularized HTEE, so that $S_{tR} \le S_{tR}^0$ and $S_{tI} \le S_{tI}^0$. 

In the left panel of figure \ref{fig2}, the real and imaginary parts of the regularized HTEE are plotted as functions of the anisotropy parameter $a$ in the limit where $a$ is arbitrary. One observes that both components of the HTEE first reach minimum values, then increase and change sign. For sufficiently large anisotropy, the parameter $a$ has opposite effects on $\widehat{S}_{tR}$ and $\widehat{S}_{tI}$. In particular, the magnitude of both components increases substantially. It is important to note that there is a point at which the effects of temperature and anisotropy cancel and $\widehat{S}_{tR}$ and $\widehat{S}_{tI}$ vanish. However, this cancellation point is not the same for the real and imaginary parts.

One may explain this behavior as follows. Increasing the anisotropy parameter 
initially breaks the symmetry of the system, disrupting the existing timelike 
correlations and reducing their magnitude. However, beyond a critical anisotropy, 
the deformation becomes strong enough to establish a qualitatively new ordered 
state, leading to a reorganization and subsequent increase in timelike correlations. 

A few comments about the holographic results and those obtained in section \ref{example} are in order.
\begin{itemize}
\item As can be seen from figures \ref{fig1} and \ref{fig2}, the imaginary parts of both the pseudo-entropy and the HTEE are decreasing functions of the anisotropy parameter. Both exhibit a minimum, then increase until reaching a point where their imaginary components vanish. Beyond this point, both imaginary parts become positive. Roughly speaking, the HTEE and pseudo-entropy exhibit similar behavior. It is worth noting that while $S_{tI}^0$ (the isotropic HTEE) is generally non-zero, the isotropic pseudo-entropy always has a vanishing imaginary part.

\item At low temperature, the HTEE increases substantially with the anisotropy parameter. %However, according to our numerical observations, the pseudo-entropy approaches zero regardless of the values of the anisotropy parameters. 

%\item Interestingly, the HTEE in a strongly coupled field theory, at least for the case considered here, selects configurations with negative anisotropy parameter, as shown in figure \ref{fig1}.

\item Previous studies have reported \cite{Jena:2024tly} that in anisotropic backgrounds of the form
\begin{equation}
ds^2 = -\frac{dt^2}{u^{2\alpha}} + \frac{1}{u^2}(du^2 + dx^2),
\end{equation}
the imaginary part of the HTEE is given by $\pi\alpha^{-1}$, a positive quantity that decreases as the anisotropy parameter $\alpha$ increases. This behavior contrasts with our results, where $S_{tI}$ initially decreases but can become positive for sufficiently large anisotropy (see figure \ref{fig2}, right panel).

This discrepancy highlights an important physical insight: the response of HTEE to anisotropy depends sensitively on {\it how} the anisotropy is introduced in the dual geometry. In the metric \eqref{10}, the anisotropy appears through the function $\mathcal{H}(u)$ multiplying the spatial $z$-direction, representing anisotropy in {\it spatial} directions at fixed temperature. By contrast, in the metric above, the anisotropy parameter $\alpha$ modifies the temporal component $g_{tt}$, which corresponds to a different kind of deformation in the dual field theory, one that effectively introduces anisotropic {\it scaling} between time and space. In our anisotropic background \eqref{metric}, increasing $a$ (or equivalently $a/T$) introduces pressure anisotropy between the $z$-direction and the transverse plane, similar to anisotropic hydrodynamics in quark-gluon plasma. This spatial anisotropy {\it suppresses} both real and imaginary parts of HTEE until a critical point, beyond which quantum correlations reorganize.
\end{itemize}

\end{document}